\documentclass[aps,prl,twocolumn,showpacs,superscriptaddress,10pt,longbibliography]{revtex4-1}
\usepackage[T1]{fontenc}

\pdfoutput=1
\usepackage[dvipsnames,rgb,dvips]{xcolor}
\usepackage{graphicx}
\usepackage{float}
\usepackage{overpic}
\usepackage{amsmath,amssymb,amsfonts}
\usepackage{hyperref}
\makeatletter
\def\amsbb{\use@mathgroup \M@U \symAMSb}
\makeatother
\usepackage{mathrsfs}
\usepackage{overpic}

\newcommand\ve[1]{\boldsymbol{#1}}

\newcommand{\ed}{\text{d}}

\newcommand{\Ku}{\ensuremath{\text{Ku}}}
\newcommand{\St}{\ensuremath{\text{St}}}

\newcommand{\epsm}{\ensuremath{{\overline{\varepsilon}}^2}}
\newcommand{\epsb}{\ensuremath{{\overline{\varepsilon}}}}

\begin{document}
 \title{Relative velocities in bidisperse turbulent suspensions}
\author{J. Meibohm}
\affiliation{Department of Physics, Gothenburg University, SE-41296 Gothenburg, Sweden}
\author{L. Pistone}
\affiliation{Department of Physics, Gothenburg University, SE-41296 Gothenburg, Sweden}
\author{K. Gustavsson}
\affiliation{Department of Physics, Gothenburg University, SE-41296 Gothenburg, Sweden}
\author{B. Mehlig}
\affiliation{Department of Physics, Gothenburg University, SE-41296 Gothenburg, Sweden}

\begin{abstract}
We investigate the distribution of relative velocities between small heavy particles of different sizes in turbulence 
by analysing a statistical model for bidisperse turbulent suspensions, containing particles with two different Stokes numbers.  
This number, $\St$, is a measure of particle inertia which in turn depends on particle size.
When the Stokes numbers are similar, the distribution exhibits power-law tails, just as in the case of equal $\St$. The power-law exponent is a non-analytic function of the mean Stokes number $\overline{\rm St}$, so that the exponent cannot be calculated in perturbation theory around the advective limit. When the 
Stokes-number difference is larger, the power law disappears, but the tails of the distribution still dominate the relative-velocity moments, if $\overline{\rm St}$ is large enough.  
\end{abstract} 
\maketitle
The dynamics of small heavy particles in turbulence plays a crucial role in many scientific problems and technological applications.
Any model of the particle dynamics must refer to the turbulence the particles experience. This is a challenge for realistic 
modelling of such systems, for their direct numerical simulation (DNS), and for experiments. Novel particle-tracking techniques and improved DNS algorithms have made
it possible to uncover striking phenomena in turbulent aerosols. 
For example, heavy particles tend to avoid the vortices of the turbulent fluid, and they form small-scale fractal patterns. Nearby particles can have very high relative velocities, an effect caused by \lq{}caustic\rq{} singularities in the particle dynamics. The analysis of statistical models
has led to substantial progress in explaining these phenomena, reviewed in Ref.~\cite{Gus16}. This analysis has offered fundamental insights about how caustics shape the distribution of relative velocities of nearby particles \cite{Sun97,Fal02,Wil05,Wil06,Bec10,Gus11b,Sal12,Bew13,Perrin2015}.

These results apply only to \lq monodisperse\rq{} suspensions of {\em identical} particles. The question is therefore
 how particles of {\em different} sizes cluster and move relative to each other. Spatial clustering of such \lq bidisperse\rq{} suspensions was analysed in
Refs.~\cite{Chu05,Bec05}. It was found that the particles cluster
onto two distinct attractors, and that the fractal distribution of separations between 
differently-sized particles is cut off at a small spatial scale, $r_{\rm c}$, that can be much larger than the particle size.

In this Letter we analyse the distribution of relative velocities of particles with different sizes.
Our results are important for the physics of turbulent aerosols, because the distribution of relative velocities determines the 
speeds of colliding particles, their collision rate, and collision outcomes \cite{Wil08,Win12,Gus14c}.
\begin{figure}
\begin{overpic}[width=\columnwidth]{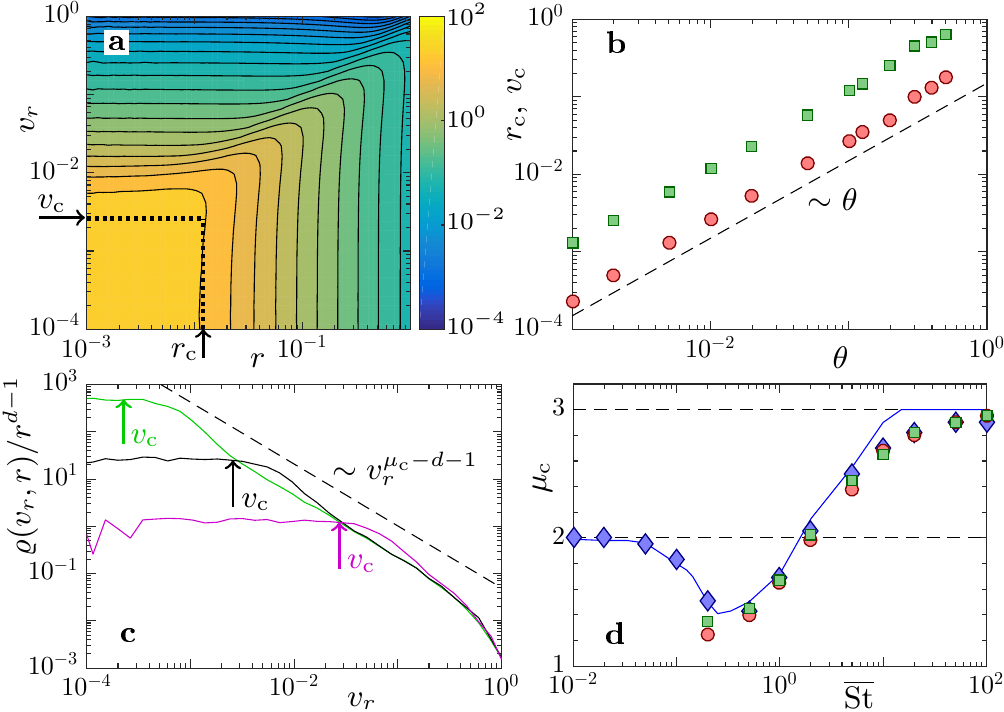}
\end{overpic}
\caption{\label{fig:attractors}
Distribution $\varrho(v_r,r)$ of relative velocities between particles with different
Stokes numbers. Statistical-model simulations in two dimensions ($d=2$) for $\Ku = 1$.
{\bf a} Contour plot of $\varrho(v_r,r)/r^{d-1}$ for $\overline{\St}=1$ and 
$\theta=0.01$ (see text).  Dotted lines show cut-off scales $r_{\rm c}$ and $v_{\rm c}$.
{\bf b} Dependence of $r_{\rm c}$ on $\theta$ (green, $\Box$),
and of $v_{\rm c}$ on $\theta$ (red, $\circ$), for $\overline{\St}=1$.
{\bf c} $\varrho(v_r,r)/r^{d-1}$ evaluated at $r<r_{\rm c}$,
for $\overline{\St}=1$ and $\theta=0.001$ (green line), $\theta=0.01$ (solid black line), and $\theta=0.1$ (magenta line).
Cross-over scales $v_{\rm c}$ (arrows).
{\bf d} Power-law exponent $\mu_{\rm c}$ versus $\overline{\St}$ for small $\theta$.
Exponent for power-law tails in $v_r$ for fixed $r$ with $\theta=0.001$ (green,$\Box$) and $\theta=0.01$ (red,$\circ$).
Exponent for power-law tails in $r$ for fixed $v_r$ with $\theta=0.01$ (blue,$\Diamond$).
Numerical data for ${\rm min}(D_2,d+1)$ where $D_2$ is the phase-space correlation dimension (solid blue line). }
\end{figure}

In a dilute
suspension of small, heavy, spherical particles the dynamics of a single particle is approximately given by Stokes law with constant $\gamma\equiv9 \rho_f \nu/(2 a^2 \rho_p)$:
\begin{align}
\label{eq:eom}
         \tfrac{\rm d}{{\rm d}t}{ \ve{{x}}} &= \ve{v}, \quad
         \tfrac{\rm d}{{\rm d}t} = \gamma [\ve{u}(\ve{x},t)-\ve{v}]\,.
\end{align}
Here $\ve x$ and $\ve v$ are particle position and velocity,
$a$ is the particle size, $\nu$ is the fluid viscosity, and
$\rho_f$ and $\rho_p$ are fluid and particle densities.
We model the turbulent fluid velocities by a random
Gaussian velocity field  $\ve u(\ve x,t)$
with zero mean, correlation time $\tau$, correlation length $\eta$, and typical
speed $u_0$, representing the universal small spatial scales of turbulence \cite{Sch14}, 
neglecting intermittent, non-Gaussian features \cite{Gus16}.
 Eq.~(\ref{eq:eom}) assumes
that the particle and shear Reynolds numbers are small. Gravitational settling \cite{Gus14e,Bec14,Ireland,Mathai2016,Parishani} is disregarded, this is valid when the turbulence is intense enough.
The dimensionless parameters of the model are the Stokes number
$\St \equiv 1/(\gamma \tau)$, a measure of particle inertia, and the Kubo number, $\Ku\equiv u_0 \tau/\eta$, measuring the persistence of the flow. 

{\em Statistics of relative velocities and separations}. Fig.~\ref{fig:attractors}{\bf a}  shows results of statistical-model simulations for $\Ku= 1$ in two spatial dimensions.
Panel {\bf a} shows the distribution $\varrho(v_r,r)$ of 
relative radial velocity $v_r$ and separation $r$ between particles with slightly different Stokes numbers.
For fixed small $v_r$, the distribution exhibits a power law as a function of $r$ for large values of $r$, and becomes uniform for small values of $r$. 
This cross over defines the cut-off scale  $r_{\rm c}$, observed and discussed in Refs.~\cite{Chu05,Bec05}.

Fig.~\ref{fig:attractors}{\bf a} shows that there is a 
second cut-off scale, $v_{\rm c}$, that distinguishes between a power law as a function of $v_r$ for large $v_r$, and a plateau for small $v_r$. 
Our numerical results demonstrate  that this new scale $v_{\rm c}$ does not depend on $r$ (Fig.~\ref{fig:attractors}{\bf a}).
This is in marked difference to the monodisperse case, where the distribution has no plateau in the limit of $r\to 0$. Panel {\bf b} 
shows that $v_{\rm c}$ depends linearly on the parameter $\theta\equiv|\St_1-\St_2|/(\St_1+\St_2)$ measuring the difference between the Stokes numbers.
Panel {\bf c} shows distributions for $r\ll r_{\rm c}$ as functions of $v_r$, 
for different values of $\theta$.  The dashed line is the power law ${|v_r|}^{\mu_{\rm c}-d-1}$,
where $d=2$ is the spatial dimension.
Panel {\bf d} demonstrates that the exponent $\mu_{\rm c}$ is approximately equal to ${\rm min}(D_2,d+1)$ where $D_2(\overline{\St})$ is the phase-space correlation dimension of a suspension 
of identical particles with Stokes number $\overline{\St}\equiv(\overline{\gamma}\tau)^{-1}$ and $\overline{\gamma} = \tfrac{1}{2}(\gamma_1+\gamma_2)$.

The shape of the distribution shown in Fig.~\ref{fig:attractors}{\bf c} has a strong effect on the moments of $v_r$.
For bidisperse suspensions, the uniform part of  $\varrho(v_r,r)$  ($r<r_{\rm c}$ and $|v_r|<v_{\rm c}$) yields an $r^{d-1}$-contribution to the moments  $\langle |v_r|^p\rangle$, just as the caustic-contribution for identical particles \cite{Gus12},
caused by uniformly distributed, uncorrelated particles with large relative velocities originating from caustic singularities \cite{Fal02,Wil05}. 
For small mean Stokes number, $\overline{\St}\ll1$, when caustics are rare, the dominant contribution to the moments of $v_r$ at small $r$ 
comes from the uniform part of the distribution. But for $\overline{\St}\sim 1$, caustics are abundant and for $p\geq 1$ the moments $\langle |v_r|^p\rangle$ are dominated by the tails of the distribution.

The uniform part of the bidisperse distribution below $r_{\rm c}$ and $v_{\rm c}$ affects the tails indirectly. Comparing the normalised mono- and bidisperse distributions we see that the tails for $\theta\neq 0$ must lie above those for $\theta=0$. This means that the moments  $\langle |v_r|^p\rangle$ must have a minimum for $\theta=0$ (at fixed $\overline{\St}$ and fixed small separation).  This is consistent with the findings of Refs.~\cite{Vol80,Miz88,Mar91,PadoanII,James}.

{\em Analysis of the white-noise limit.}
We now show that all of the above observations can be explained qualitatively by analysing
a one-dimensional white-noise model.
We linearise Eq.~(\ref{eq:eom}), 
$\Delta  u(x,t)\sim  A(t) \Delta x$,
to obtain equations for small separations $\Delta x\equiv x_2-x_1$.
The spatial dependence in $A(x,t)\equiv$ $\partial_x u(x,t)$ is neglected here,
this disregards preferential sampling which is not important in the white-noise limit \cite{Gus16}.
The gradient $A(t)$ has zero mean and correlation 
function $ \langle A(t_1) A(t_2) \rangle = 3(u_0/\eta)^2\exp( -|t_2 - t_1|/\tau)$.
For $\theta \neq 0$ there is an additional stochastic driving,  $u(x_1,t)+u(x_2,t) \approx 2u(\overline{x},t)$
where $\overline{x} \equiv \tfrac{1}{2}(x_1+x_2)$.
Neglecting the spatial dependence we write $B(t)\equiv 2u(\overline{x},t)$. The two noise terms $A(t)$ and $B(t)$ are uncorrelated, and $\langle B(t_1)B(t_2)\rangle = 4u_0^2\exp(-{|t_2 - t_1|}/{\tau})$. We use the de-dimensionalisation
$\tilde t \equiv t\overline{\gamma}\,, \tilde{ x} \equiv x/\eta\,,
\tilde{ v} \equiv  v/(\eta\overline{\gamma})\,, \tilde{ u} \equiv  u/(\eta\overline{\gamma})$.
Dropping the tildes we find:
\begin{equation}\label{eq:B}\hspace*{-2mm}
\tfrac{\rm d}{{\rm d}t} \Delta x \!=\!\Delta v \quad\!\!\!\mbox{and}\!\!\!\quad
\tfrac{\ed}{\ed t}\begin{bmatrix}\Delta v	\\ 2\overline{v}	\end{bmatrix} =
\begin{bmatrix}1&\theta\\ \theta&1\end{bmatrix}\begin{bmatrix}A(t) \Delta x\!-\!\Delta v\\ B(t)\!-\!2\overline{v}
\end{bmatrix}\,,
\end{equation}
where $\Delta v \equiv v_2-v_1$ is the relative velocity.
The white-noise limit is taken by letting $\overline{\St}\to\infty$ and $\Ku\to0$ such that $\overline{\varepsilon}^2\equiv 3\Ku^2\,\overline{\St}$ stays finite. In this limit we obtain 
$\langle A(t_1) A(t_2)	\rangle	=   \tfrac{3}{4} \langle B(t_1) B(t_2)\rangle=2\epsm\delta(t_2-t_1)$.
The white-noise problem has three coupled dynamical variables, $\Delta x$ and $\Delta v$
as in the monodisperse case, and the mean velocity $\overline{v} \equiv{\tfrac{1}{2}}(v_1+v_2)$. 
We write $\varrho(\Delta v, \Delta x) \equiv \int {\rm d}\overline{v}\,P(\Delta v,\Delta x,\bar v)$,
where $P(\Delta v,\Delta x,\bar v)$ is the steady-state distribution of $\Delta v,\Delta x$ and $\bar v$.

{\em Cross-over scale $v_{\rm c}$.}
For small values of $\theta$, the mean velocity $\overline{v}$ is approximately governed by an Ornstein-Uhlenbeck process driven by $B$.  In  the equation for $\Delta v$ we identify three competing terms: the two noise terms $V_1\equiv A\Delta x$ and $V_2\equiv \theta(B-2\overline{v})$, and the damping $-\Delta v$.

\begin{figure}[t]
\begin{overpic}[width=\linewidth]{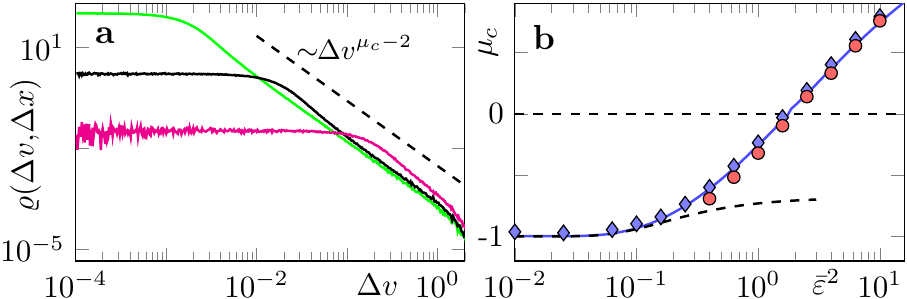}
\end{overpic}
\caption{\label{fig:whitenoise}
White-noise results for distribution of relative velocities
for two different Stokes numbers, one-dimensional statistical model. {\bf a} Distribution of $\Delta v$ at small $\Delta x$ for $\epsm=2$.
{\bf b} Exponent $\mu_{\rm c}$ as a function
of $\varepsilon$ from white-noise simulations of the two-Stokes model for $\theta=0.01$ (symbols), obtained from fit to 
$\Delta x$-tails (blue,$\Diamond$),
and from fit to $\Delta v$-tails (red, $\circ$).
Result for equal $\St$, solid line,
obtained from numerical solution of Eq.~(\ref{eq:sep2}). 
Also shown is Eq.~(\ref{eq:na}), dashed line.}
\end{figure}

For small relative velocities ($|\Delta v| \to 0$), the damping term is negligible. In this limit, the nature of the dynamics
depends on whether $|\Delta x| \ll \theta$ or $|\Delta x| \gg \theta$. In the first case
$V_2$ dominates over $V_1$, in the second case $V_1$ dominates.  We find the spatial cut off $r_{\rm c}$ by estimating the value of $\Delta x$ where $V_1$ and $V_2$ are of the same order.
This yields $r_{\rm c}\propto\sqrt{\Phi_{V_2}/\Phi_A} \propto \theta$, where $\Phi_X$ is the integrated correlation function of $X$.
Now consider small separations ($|\Delta x|\ll r_{\rm c}$) and finite $\Delta v$. In this case $V_1$ is negligible. Now we must consider whether
$|\Delta v|\ll\theta$ or $|\Delta v|\gg\theta$. In the first case the noise term $V_2$ dominates, so that
the dynamics is diffusive. When $|\Delta v|\gg\theta$, on the other hand, the dynamics is
deterministic, and the noise plays no role.  Between these two regimes we find the
cut-off scale $v_c\propto \sqrt{\Phi_{V_2}}\propto\theta$.
These considerations explain the linear $\theta$-dependencies of the cutoff-scales $r_{\rm c}$ and $v_{\rm c}$,
Figs.~\ref{fig:attractors}{\bf c} and Fig.~\ref{fig:whitenoise}{\bf a}.
Finally, consider the case where $V_2$ is negligible ($|\Delta x|\gg r_{\rm c}$ or  $|\Delta v|\gg v_{\rm c}$).
In this limit the dynamics is essentially that of monodisperse suspensions,
but with mean Stokes number  $\overline{\St}$.
For $|\Delta x|>r_{\rm c}$ or $|\Delta v|>v_{\rm c}$ the distribution is expected to show the same power-law tails as the
monodisperse case (Fig.~\ref{fig:attractors}{\bf c}).

{\em Power-law tails.} 
We now discuss how the power-law tails observed in bidisperse suspensions for small $\theta$ are related
to the power laws observed in the monodisperse case. 
For $\theta\ll1$ we decompose the joint distribution $P(\Delta v,\Delta x,\bar v)=\varrho_0(\Delta v, \Delta x) p(\overline{v}) + \delta P(\Delta v, \Delta x, \overline{v})$, where $\varrho_0(\Delta v, \Delta x)$ equals the distribution for monodisperse particles ($\theta=0)$.
It follows from the analysis above that the relative dynamics is diffusive for $|\Delta x|\ll r_{\rm c}$ and $|\Delta v|\ll v_{\rm c}$. In this case $\delta P(\Delta v,\Delta x,\bar v)$ cannot be neglected. 
For $|\Delta x|\gg r_{\rm c}$ or $|\Delta v|\gg v_{\rm c}$, by contrast, 
the dynamics is that of the monodisperse suspension and
$\varrho_0(\Delta v, \Delta x)$ dominates over $\delta P(\Delta v,\Delta x,\bar v)$.  
To study the tails it thus suffices to consider $\varrho_0(\Delta v, \Delta x)$.
We follow Ref.~\cite{Gus11b} and make the separation ansatz:
$\varrho_0(\Delta x,z) = \sum_\mu A_\mu g_\mu(\Delta x) Z_\mu(z)$ with $z\equiv\Delta v/\Delta x$ .
Inserting this ansatz into the Fokker-Planck equation corresponding to Eq.~\eqref{eq:B} we obtain
\begin{equation}
\label{eq:sep2}
g_\mu(\Delta x)\!=\!|\Delta x|^{\mu-1}\mbox{and}\,\,\, \tfrac{\rm d}{{\rm d}z}(z\!+\!z^2\!+\!\epsm \tfrac{\rm d}{{\rm d}z})Z_\mu\!=\!\mu z Z_\mu
\end{equation}
to lowest order in $\theta$. This is the equation for a suspension of identical particles \cite{Gus11b} with Stokes number $\overline{\St}$. In a slightly different form, Eq.~(\ref{eq:sep2}) was used to model the effect of wall collisions in turbulent suspensions \cite{Bel14}.

Particle-exchange symmetry \cite{Gus11b} requires that $Z_\mu(z)$ is symmetric for large $|z|$,
$\lim_{z\to\infty}Z_{\mu}(z)/Z_{\mu}(-z)=1$. Numerical analysis shows that Eq.~(\ref{eq:sep2}) exhibits a discrete set of such solutions \cite{Gus14c}. For small $\overline{\varepsilon}$ only two values are allowed: $\mu=0$ and $\mu=\mu_{\rm c}(\overline{\varepsilon})$. 
At large values of $|z|$ one finds that $Z_{\mu_{\rm c}}(z)\sim |z|^{\mu_{\rm c}-2}$.
In the monodisperse case, $\mu_{\rm c}(\epsb)$ equals $D_2$ 
for $\overline{\varepsilon} > \overline{\varepsilon}_{\rm c}$.
For $\overline{\varepsilon} < \overline{\varepsilon}_{\rm c}$, by contrast, 
particle paths coalesce exponentially \cite{Wil03}, there is no power law steady-state, and $\mu_{\rm c}(\epsb)$ becomes negative and loses its meaning as correlation dimension. The distribution $g_{\mu_{\rm c}}(\Delta x)=|\Delta x|^{\mu_{\rm c}-1}$ is not normalisable 
for $\mu_{\rm c}\leq 0$.

In the bidisperse case, this divergence is regularised since diffusion dominates for $|\Delta x|\ll r_{\rm c}$ and
$|\Delta v|\ll v_{\rm c}$.  More precisely, $\delta P(\Delta v,\Delta x,\bar v)$ regularises $P(\Delta v, \Delta x, \overline{v})$. Thus a power-law steady-state
distribution is obtained, even for $\overline{\varepsilon} < \overline{\varepsilon}_{\rm c}$.
This is analogous to the effect of small-scale diffusion that regularises a power-law steady
state distribution of separations \cite{Wil15}.

The full solution of (\ref{eq:sep2}) consists of a sum of two terms,
 $A_0 \Delta x^{-1} Z_0(z) + A_{\mu_{\rm c}}  |\Delta x|^{\mu_{\rm c}-1} Z_{\mu_{\rm c}}(z)$.
As was discussed in \cite{Gus11b}, the separation ansatz for $\varrho(\Delta x,z)$ is strictly valid only for $|\Delta x|\ll1$. The $A_0$-term is therefore subdominant for $\mu_{\rm c}<0$. For $\mu_{\rm c}>0$ 
we expect $A_0\sim\theta\ll A_{\mu_{\rm c}}$ since $A_0$ must vanish in order for the distribution to be normalisable over $\Delta x=0$ as $\theta\to 0$, consistent with the numerical small-$\theta$ results in Fig.~\ref{fig:whitenoise}{\bf a}. 
Therefore we disregard the $\mu=0$-term for all values of $\overline{\varepsilon}$.

Fig.~\ref{fig:whitenoise}{\bf b} shows that for $\theta\ll1$ the exponents of both the $\Delta x$- and 
$\Delta v$- distributions (symbols) follow the numerically calculated $\mu_{c}(\bar\varepsilon)$ very precisely (solid line).
 The asymptote $Z_{\mu_{\rm c}}(z)\sim |z|^{\mu_{\rm c}-2}$ for large $|z|$ 
gives rise to the power-law tails in the relative-velocity distribution of the form $|\Delta v|^{\mu_{\rm c}-2}$. In $d$ spatial dimensions
a similar argument yields tails on the form $v_{r}^{\mu_{\rm c}-d-1}$. This explains the power laws observed
in Figs.~\ref{fig:attractors}{\bf c} and \ref{fig:whitenoise}{\bf a}. 

In summary the white-noise analysis qualitatively explains not only the existence of the plateau in the relative-velocity distribution and the corresponding cutoff scales $r_c$ and $v_c$. It also yields 
the linear dependence of these scales upon $\theta$, and explains the power-law tails. Thus,
the one-dimensional white-noise analysis qualitatively explains all the observed features in Fig.~\ref{fig:attractors}.

{\em Failure of perturbation theory in $\overline{\varepsilon}$.}
Fig.~\ref{fig:whitenoise}{\bf b} demonstrates that the exponent $\mu_{\rm c}$      
depends very sensitively on $\epsb$, for small values of $\epsb$. Here we show 
that this dependence is non-analytic. This is important, because it means
that perturbation theory in $\epsb$ fails. This is, in turn, a likely reason \cite{Wil14} why Borel resummation of perturbation theory does not yield accurate results for the correlation dimension $D_2$ (Fig.~1 in Ref.~\cite{Wil10b}). There is to date no theory explaining these observations.

The exponent $\mu_{\rm c}$ appears as a generalised eigenvalue in Eq.~(\ref{eq:sep2}), but exact
solutions to Eq.~(\ref{eq:sep2}) are known only for $\mu = 0$ \cite{Wil03},
$\mu=-1$ \cite{Der07}, and in the large $\overline{\varepsilon}$-limit \cite{Scho02}. 
The physical solution at small values of $\epsb$ corresponds to
the eigenvalue $\mu_{\rm c}(\epsb)$. The solution for $\mu=-1$ is unphysical since it does not obey particle-exchange symmetry.

Numerical analysis reveals that $\mu_{\rm c}(\epsb)\to -1$ as $\epsb\to 0$.
We write $\mu_{\rm c}\!=\!-1\!+\!\delta\mu$, where $\delta\mu\geq 0$ is an \lq eigenvalue splitting\rq{} that vanishes as $\overline{\varepsilon}\to 0$, and attempt to find the physical solution by perturbation theory in $\delta\mu$:
\begin{align}
\label{eq:expand}
	Z_{\mu_{\rm c}} (z) = Z^{(0)} (z) + \delta\mu\,Z^{(1)}(z) + \delta\mu^2 Z^{(2)}(z) +\ldots.
\end{align}
Substituting (\ref{eq:expand}) into \eqref{eq:sep2} we obtain a hierarchy of differential equations for $Z^{(n)}$ (Supplemental Material \cite{supp}).
The lowest-order solution is the unphysical one, $Z^{(0)} = Z_{-1}$. Its general form is known \cite{Der07}:
\begin{equation} \label{eq:Zsol0}
	Z_{-1} \!=\! C_1 (z\!+\!1){\rm e}^{-U(z)} \!\!+C_2(z\!+\!1) \! \int_{-\infty}^z\!\!\!\!\!\ed t\,\frac{{\rm e}^{U(t)-U(z)}}{(t+1)^2}\,.
\end{equation}
Here  $U(z)\equiv (\tfrac{1}{3}z^3+\tfrac{1}{2}z^2)/\epsm$, and $C_1$ and $C_2$ are constants. 
The first-order solution reads:
\begin{equation}
	Z^{(1)}\!=\!\frac{z\!+\!1}{\epsm}\!\!
\int_{-\infty}^z\!\!\!\!\!\!\ed t\,\frac{{\rm e}^{U(t)-U(z)}}{(t+1)^2}\int_{-1}^t\!\!\!\!\ed t' t'(t'\!+\!1) Z_{-1}(t')\,. \label{eq:Zsol1}
\end{equation}
The integrand involves $Z_{-1}$, and we must take $C_1=0$ in Eq.~(\ref{eq:Zsol0}) for $Z_{-1}$ since
the first term in (\ref{eq:Zsol0}) is not normalisable. We further choose  $C_2=-\epsm$ because it normalises the large-$z$ tails of $Z_{-1}$ to $Z_{-1}\sim -z^{-3}$.
To evaluate (\ref{eq:Zsol1}) we use the fact that $\delta\mu\to 0$ as $\overline{\varepsilon}\to 0$ and apply a WKB approximation \cite{Ben78} of $Z_{-1}$ that becomes exact in the limit $\overline{\varepsilon}\to 0$. The equation for $Z_{-1}$ has two
\lq turning points\rq{} at $z=-1,0$ where the WKB approximation breaks down.
We find locally exact solutions and use these to match the WKB solutions across the turning points.
Details of this asymptotic-matching method are given in the Supplemental Material \cite{supp}.
In this way we find $Z^{(1)} \sim  |z|^{-3}\log |z|$ for large
negative values of $z$, and 
$Z^{(1)} \sim |z|^{-3}[2\pi {\rm e}^{1/(6\epsm)} - \log |z|]$ 
for large positive $z$.
Imposing particle-exchange symmetry yields
\begin{equation}
\label{eq:na}
\mu_{\rm c} = -1+\delta\mu\quad\mbox{with}\quad \delta\mu = {\pi}^{-1} {\rm exp}[{-1/(6\epsm]}\,.
\end{equation}
This result is shown as a dashed line in Fig.~\ref{fig:whitenoise}{\bf b}. It is in good agreement with the numerical data for small $\epsb$.

We emphasise that the non-analytic dependence (\ref{eq:na}) cannot be obtained
by the $\epsb$-perturbation theory described in Refs.~\cite{Meh04,Dun05,Gus16}. That approach starts by rescaling $z\to \overline{\varepsilon} z$, and solves the resulting equation
$\tfrac{\rm d}{{\rm d}z}(z+\overline{\varepsilon} z^2+\tfrac{\rm d}{{\rm d}z})Z_\mu\!= \overline{\varepsilon} \mu z Z_\mu$ perturbatively in $\epsb$. 
The solution remains essentially Gaussian. 
Even for very small $\epsb$, however, the $z$-dynamics can escape from $z\approx 0$ to $-\infty$ by forming caustics \cite{Gus16}. This gives rise to power-law tails that are not described by 
the local theory. Higher orders in $\epsb$ improve the accuracy only locally, inside an $\epsb$-sized \lq{}boundary layer\rq{} \cite{Hinch} around $z=0$, but not in the tails. Consequently, perturbation theory fails to give the non-analytical dependence (\ref{eq:na}).

Our theory, by contrast, provides a uniform approximation valid for small $\overline{\varepsilon}$ and arbitrary $z$. 
 It allows to compute the exponentially small eigenvalue splitting $\delta\mu$ because
it captures the formation of caustics. For small $\epsb$, the rate $J$ of caustic formation is $J\sim (2\pi)^{-1} {\rm e}^{-1/6\epsm}$ \cite{Gus16}. It follows that $\mu_{\rm c} \sim 2J - 1$ at small values of $\epsb$.

We now show  that caustics cause similar problems in small-$\St$ expansions \cite{Max87,Elp96,Bal01,Chu05,Wil07,Gus16}.
In one dimension, the equation for $z$ reads in dimensional units
\begin{equation}
\label{eq:pers} \tfrac{{\rm d}}{{\rm d}t} z = -\gamma z-z^2 + \gamma A(x,t)\,.
\end{equation}
The Stokes number used in DNS of heavy particles in turbulence \cite{Bec06,Bec07} is $\kappa \equiv u_0/(\eta\gamma)$ \cite{Gus16}.  
Therefore we consider the small-$\kappa$ limit of (\ref{eq:pers}), taking $A$ to be constant (in
this limit the local fluid-velocity gradients are roughly constant during the time $\eta/u_0$ \cite{Gus16}). As $\kappa\to0$ we find that $z\sim A$, 
with distribution $P(A) = (6\pi u_0^2/\eta^2)^{-1/2} \exp[-A^2/(6u_0^2/\eta^2)]$, localised around zero.  As in the white-noise case, this local approximation fails when caustics allow $z$ to escape to $-\infty$. Kramers-escape theory for coloured noise \cite{Tsi88,Kan89} reveals that caustics occur in the persistent limit when $A<-\gamma/4$, with probability $p\! =\!\int_{-\infty}^{-\gamma/4}\!\!\ed A\,P(A)$. Evaluating this integral for $\kappa \ll 1$, one obtains a non-analytic dependence,  $p \propto \exp[-1/(96 \kappa^2)]$, consistent with 
Ref.~\cite{Gus13a}. In DNS a similar non-analytic dependence is found \cite{Fal07c}, albeit of a slightly 
different form because the turbulent velocity-gradients are non-Gaussian. We conclude that the perturbation theories in the white-noise limit and in the persistent limit fail for the same reason. Both expansions are valid only near $z\!=\!0$ and fail in the presence of caustics. 

{\em Conclusions.}
We analysed the distribution of relative velocities in turbulence between small, heavy particles with different Stokes numbers $\St$.
We demonstrated that the difference in $\St$ causes diffusive relative motion at small separations, giving rise to a plateau in the distribution for relative velocities smaller than a cut off $v_{\rm c}$. This is in qualitative agreement with DNS \cite{PadoanIII}. Fig.~2 in Ref.~\cite{PadoanIII} shows the $v_r$-distribution for a bidisperse suspension.  
The cut off $v_{\rm c}$ depends linearly on $\theta$, the parameter characterising the difference in Stokes numbers.

At small values of $\theta$
the distribution exhibits
algebraic tails $|v_r|^{\mu_{\rm c}-d-1}$ as in the monodisperse case. The exponent $\mu_{\rm c}$ is determined 
by the phase-space correlation dimension of a monodisperse system with mean Stokes number $\overline{\St}$. When $\overline{\St}$ is $\mathcal{O}(1)$ or larger, the moments of relative velocities at small separations are dominated by the tails of the distribution. 
At small $\theta$ the tails are power laws, but as $\theta$ becomes larger, our statistical-model simulations show that the tails gradually disappear, consistent with $v_{\rm c}\propto\theta$.
The DNS of Ref.~\cite{PadoanIII} are for quite large $\theta$. It would be of interest to perform DNS for small values of $\theta$ to find the power-law tails we predict here. 

Our analysis shows how sensitive the distribution is to polydispersity. This is important for
experiments tracking the dynamics of $\mu$m-size particles in turbulence \cite{Bew13}, where strict monodispersity is difficult to achieve.

We explained these observations using a one-di\-men\-si\-onal statistical model. 
The difference in Stokes numbers regularises the distribution, so that 
the bidisperse model has a power-law steady-state at small $\overline{\varepsilon}$. This makes it possible
to compute how the power-law exponent $\mu_{\rm c}$ depends on the inertia parameter $\overline{\varepsilon}$. 
We demonstrated that the dependence of $\mu_{\rm c}$ upon $\overline{\varepsilon}$ is not analytic.

Our theory explains why small-$\epsb$ expansions fail to give non-analytic contributions to physical quantitites
in the white-noise limit.  It is likely that non-analytic terms in white-noise expansions for the correlation dimension \cite{Wil14} and Lyapunov exponents \cite{Meh04,Dun05} in two and three dimensions have similar origins, and we speculate that Eq.~(\ref{eq:na}) contains only the first two terms in an infinite series of the form  \cite{Dor14} $\sum_{k=0}^\infty {\rm e}^{-k/(6\epsm)} \sum_{m=0}^\infty b_m^{(k)} \epsb^{2m}$. There could
also be logarithmic terms, $\log^n\epsm$.

More generally we explained that small-$\St$ expansions for heavy particles in turbulence \cite{Max87,Elp96,Bal01,Chu05,Wil07,Gus16} suffer from similar problems when caustics occur. This indicates that matched asymptotic expansions (or similar methods) are required to explain the characteristic minimum \cite{Bec07} of the correlation dimension as a function of $\St$.  

Our predictions can be directly tested by experiments or by DNS
of turbulent bidisperse turbulent suspensions. We note, however, that our analysis pertains
 to the dynamics in the dissipative range. At higher Stokes numbers,
 when separations between particle pairs explore the inertial range \cite{Meh07,Gus08b},
we expect corrections to the power-law exponents derived here.

\eject   
\begin{acknowledgments} 
This work was supported by Vetenskapsr\aa{}det [grant number 2013-3992], Formas [grant number 2014-585], and by the grant \lq Bottlenecks for particle growth in turbulent aerosols\rq{} from the Knut and Alice Wallenberg Foundation, Dnr. KAW 2014.0048.  The numerical computations used resources provided by C3SE and SNIC.
\end{acknowledgments}

\end{document}


\title{Supplemental material for \lq Relative velocities in bidisperse turbulent suspensions\rq{}}
\author{J. Meibohm}
\affiliation{Department of Physics, Gothenburg University, SE-41296 Gothenburg, Sweden}
\author{L. Pistone}
\affiliation{Department of Physics, Gothenburg University, SE-41296 Gothenburg, Sweden}
\author{K. Gustavsson}
\affiliation{Department of Physics, Gothenburg University, SE-41296 Gothenburg, Sweden}
\author{B. Mehlig}
\affiliation{Department of Physics, Gothenburg University, SE-41296 Gothenburg, Sweden}

\maketitle
This Supplemental Material gives additional details on the derivation of Eq.~(7) in the main text, describes how to obtain the WKB solution for $Z_{-1}(z)$ that is used to evaluate Eq.~(7), and how to obtain Eq.~(8).

\section{Derivation of Eq.~(7)}
\label{sec:Z1}

Substituting $\mu= -1+\delta\mu$ into the differential equation Eq.~(4) in the main text we find:
\begin{align}\label{eq:sep2}
	\frac{\ed}{\ed z}\left(z+z^2 + \epsm\frac{\ed}{\ed z}\right)Z_\mu(z)	+zZ_\mu(z) = \delta\mu\, z Z_\mu(z)
\end{align}
This equation is solved by a perturbation expansion in the small parameter $\delta\mu$. We make the ansatz
\begin{align}
\label{eq:expand}
	Z_{\mu_{\rm c}} (z) = Z^{(0)} (z) + \delta\mu\,Z^{(1)}(z) + \delta\mu^2 Z^{(2)}(z) +\ldots.
\end{align}
Substituting Eq.~(\ref{eq:expand}) into Eq.~\eqref{eq:sep2} we obtain a hierarchy of differential equations for $Z^{(n)}$:
\begin{align}
	\frac{\ed}{\ed z}\left(z+z^2 + \epsm\frac{\ed}{\ed z}\right)Z^{(0)}_\mu(z)	+zZ^{(0)}_\mu(z) &= 0\,,	\label{eq:order0}\\
	\frac{\ed}{\ed z}\left(z+z^2 + \epsm\frac{\ed}{\ed z}\right)Z^{(n)}_\mu(z)	+zZ^{(n)}_\mu(z) &= z Z^{(n-1)}_\mu(z)\,,\quad n\geq 1\,.\label{eq:ordern}
\end{align}
Eq.~\eqref{eq:order0} is just Eq.~(\ref{eq:sep2}) for $\delta\mu=0$, for which the two linearly independent solutions are known \cite{Der07}. 
We call these solutions $Z^{(a)}_{-1}$ and $Z_{-1}$:
\begin{align}
	Z^{(a)}_{-1}(z) 	&= (z\!+\!1){\rm e}^{-U(z)}\,,	\label{eq:Za}\\
	Z_{-1}(z)	&=	-\frac{(z+1)}{\epsm}\int_{-\infty}^z\!\!\!\!\!\ed t\,\frac{{\rm e}^{U(t)-U(z)}}{(t+1)^2}\,, \label{eq:Zb}
\end{align}
with   $U(z)\equiv (\tfrac{1}{3}z^3+\tfrac{1}{2}z^2)/\epsm$. Since $Z^{(a)}_{-1}(z)$ diverges exponentially for $z\to-\infty$ it is not normalisable.
Thus it is not an admissible solution to Eq.~\eqref{eq:order0}. We therefore need to consider $Z_{-1}(z)$ as the inhomogeneity on the right-hand side of equation \eqref{eq:ordern} for $n=1$:
\begin{align}
	\frac{\ed}{\ed z}\left(z+z^2 + \epsm\frac{\ed}{\ed z}\right)Z^{(1)}_\mu(z)	+zZ^{(1)}_\mu(z) = z Z_{-1}(z)\,,\quad n\geq 1\label{eq:order1}\,.
\end{align}
We solve this equation by the method of reduction of order \cite{Ben78}. We make the ansatz
\begin{align}
	Z^{(1)}(z) = Z^{(a)}_{-1}(z) f(z)
\end{align}
and derive a first-order differential equation that involves the unknown function $f(z)$. The differential equation is solved by integration. We obtain
\begin{equation}
	Z^{(1)}(z)\!=\!\frac{z\!+\!1}{\epsm}\!\!
\int_{-\infty}^z\!\!\!\!\!\!\ed t\,\frac{{\rm e}^{U(t)-U(z)}}{(t+1)^2}\int_{-1}^t\!\!\!\!\ed t' t'(t'\!+\!1) Z_{-1}(t')\,. \label{eq:Z1}
\end{equation}
This is Eq.~(7) in the main text.  Evaluating this expression further is difficult. 
In the limit $\epsb\to0$, however, we can use a Wentzel-Kramers-Brillouin 
(WKB) approximation \cite{Ben78} to evaluate the integral.

\section{WKB approximation of $Z_{-1}(z)$}
\label{sec:wkb}
The WKB method allows to obtain global approximations to linear differential equations where the highest derivative is multiplied by a small expansion parameter. In Eq.~\eqref{eq:order0} the small parameter is $\epsb$. 
One starts with the ansatz
\begin{align}
	Z_{-1}(z) \sim C\, {\rm e}^{S(z)/\epsm}\,,
\end{align}
where $C$ is a constant. By inserting this ansatz into Eq.~\eqref{eq:order0} one obtains
\begin{align}
	\epsb^2\left[S''(z)+(3z+1)\right] + (z+z^2)S'(z) + S'(z)^2 = 0\,,
\end{align}
where primes denote derivatives w.r.t $z$.
Up to this point the procedure is still exact. 
Now one expands $S$ in $\overline{\varepsilon}^2$:
\begin{equation}
S(z)=S^{(0)}(z)+\epsm S^{(1)}(z)+\overline{\varepsilon}^4 S^{(2)}(z)+\ldots,
\end{equation}
and evaluates the resulting equation order by order in $\overline{\varepsilon}$.
In this way one finds a hierarchy of differential
equations for $S^{(n)}$:
\begin{align}
	[S^{(0)'}]^2 &= -(z+z^2)S^{(0)'}						\label{eq:WKB0}\\
	S^{(1)'}	&=	-\frac{S^{(0)''}+(3z+1)}{z+z^2+ 2S^{(0)'}}	\label{eq:WKB1}\\
	 S^{(n)'} 	&=	-\frac{S^{(n-1)''}+\sum_{k=1}^{n-1} S^{(n-k)'}S^{(k)'}}{z+z^2+2S^{(0)'}}	\,,	\qquad n\geq2\,.
\end{align}
We solve the lowest two orders, Eqs.~\eqref{eq:WKB0} and \eqref{eq:WKB1},
This yields the following approximation
for $Z_{-1}(z)$ as $\epsb\to0$:
\begin{align}\label{eq:WKB}
	Z^{\rm WKB}_{-1}(z)	\sim	C_1\, (z+1)\exp\left[ -U(z)\right] + \frac{C_2}{z(z+1)^{2}}\left[1+\epsm \left(\frac{1}{z^2 (z+1)^2}+\frac{4}{z (z+1)^2}\right)\right]\,.
\end{align}
We see that the first term in Eq.~\eqref{eq:WKB} is equal to $Z^{(a)}_{-1}$, one of the exact solutions to the 
differential equation \eqref{eq:order0}. 
This means that the WKB expansion for this part of the solution truncates after two terms. The WKB approximation to the second term, on the other hand, does not truncate and increasing orders become more and more complicated. This term also suffers from divergences at the \lq turning points\rq{} $z=0$ and $z=-1$. 
We must cure these divergences by performing asymptotic matching across the turning points. In the present case we require a uniform approximation to the differential equation to order $\mathcal{O}(\epsm)$.

To find the solutions close to the turning points, we follow
\cite{Ben78} and express the differential equation for $Z_{-1}$ in terms of the \lq{}turning-point coordinates\rq{} $t=(z+1)/\epsb$ and $s=z/\epsb$ for the turning points at $z=-1$ and $z=0$, respectively. This gives 
\begin{align}\label{eq:TPl}
	(3\epsb \, t - 2)Z_{-1}(z) + (\epsb \, t^2 - t)Z_{-1}(z) + Z''_{-1}(t) = 0\,,
\end{align}
and
\begin{align}\label{eq:TPr}
	(3\epsb \, s + 1)Z_{-1}(s) + (\epsb \, s^2 + s)Z_{-1}(s) + Z''_{-1}(s) = 0\,.
\end{align}
These second-order equations have two solutions each.  One solution for
each of these equations is already known since the first of our WKB solutions is exact. 
In terms of the turning-point coordinates $t$ and $s$, 
the exact solution $Z^{(a)}_{-1}$ reads
\begin{align}
	Z^{(a)}_{-1}(t) = C^{(l)}_1 \,\epsb\, t\, {\rm e}^{-\frac1{6\epsm}}\exp\left[\frac{t^2}{2}-\epsb\,\frac{t^3}{3}\right]\sim C^{(l)}_1 \,\epsb\, t\, {\rm e}^{-\frac1{6\epsm}}{\rm e}^{\frac{t^2}{2}}\left(1-\epsb\,\frac{t^3}{3}\right)
\end{align}
and
\begin{align}
	Z^{(a)}_{-1}(s) = C^{(r)}_1 \,\left(\epsb s +1\right)\exp\left[-\frac{s^2}{2}-\epsb\,\frac{s^3}{3}\right]\sim C^{(r)}_1 \, {\rm e}^{-\frac{s^2}{2}}\left(1+\epsb\,s-\epsb\,\frac{s^3}{3}\right)\,,
\end{align}
respectively, where we expanded to first order in $\epsb$ around both turning points.
To obtain the corresponding approximations to the remaining solutions of the two equations \eqref{eq:TPl} and \eqref{eq:TPr} is less straightforward. We first take the zeroth-order approximations to Eqs.~\eqref{eq:TPl} and \eqref{eq:TPr} and obtain
\begin{align}
	Z_{-1}''(t)-t Z_{-1}'(t)		&\sim	2Z_{-1}(t)\,, 
\end{align}
and
\begin{align}
	Z_{-1}''(s)+s Z_{-1}'(s)	&\sim	-Z_{-1}(s)\,,
\end{align}
respectively. These equations must be solved to find the lowest-order solutions
\begin{align}
	Z_{-1}(t) &=  Z^{(a)}_{-1}(t)+ C^{(l)}_2\left[1+\sqrt{\frac{\pi}{2} }{\rm e}^{\frac{t^2}{2}} t \,\text{erf}\left(\frac{t}{\sqrt{2}}\right)\right]+\mathcal{O}(\epsilon) \label{eq:tpL}
\end{align}
and
\begin{align}
	Z_{-1}(s) &=  Z^{(a)}_{-1}(s) +C^{(r)}_2\sqrt{\frac{\pi }{2}} {\rm e}^{-\frac{s^2}{2}} \text{erfi}\left(\frac{s}{\sqrt{2}}\right)+\mathcal{O}(\epsilon)\,,
\end{align}
near the turning points at $z=-1$ and $z=0$, respectively. The next order in $\epsb$ is obtained by making the following ansatz for $Z_{-1}$ around the turning points
\begin{align}
	Z_{-1}(t) &\sim	 Z^{(a)}_{-1}(t) +C^{(l)}_2\left[1+\epsb\, f_l(t)+\left[1+ \epsb\,g_l(t)\right]\sqrt{\frac{\pi}{2} } t \,{\rm e}^{\frac{t^2}{2}}\text{erf}\left(\frac{t}{\sqrt{2}}\right)\right]\,,	\\
	Z_{-1}(s) &\sim  Z^{(a)}_{-1}(s) +C^{(r)}_2\left[\epsb\,f_r(s)+\left[1+\epsb\, g_r(s)\right]\sqrt{\frac{\pi }{2}} {\rm e}^{-\frac{s^2}{2}} \text{erfi}\left(\frac{s}{\sqrt{2}}\right)\right]\,,	\\
\end{align}
with
\begin{align}
	f_i(x)=\sum_{n=0}^{\infty} a_i^{(n)} x^n\,,	\qquad	g_i(x)=\sum_{n=0}^{\infty} b_i^{(n)} x^n\,,
\end{align}
and $i=l,r$. Inserting these expressions into the differential equations \eqref{eq:TPl} and \eqref{eq:TPr}, respectively and evaluating to first order in \epsb{} allows to fix the constants to
\begin{align}
	a^{(1)}_l &= \frac13\,,	\qquad	a^{(3)}_l = -\frac13\,,	\qquad b^{(3)}_l = -\frac13	\,,	\\
	a^{(0)}_r &= -\frac83\,,	\qquad	a^{(2)}_r = \frac13\,,	\qquad b^{(1)}_r =1\,, \qquad b^{(3)}_r =-\frac13\,.
\end{align}
The remaining constants are found to be zero.
To first order in \epsb{} the solutions around the turning points now read
\begin{align}
	Z_{-1}(t) &\sim	C^{(l)}_1 \,\epsb\, t\, {\rm e}^{-\frac1{6\epsm}}{\rm e}^{\frac{t^2}{2}}\left(1-\epsb\,\frac{t^3}{3}\right)+C^{(l)}_2\left[1+\epsb\left(\frac{t}{3}-\frac{t^3}{3}\right)+\sqrt{\frac{\pi }{2}}t\, {\rm e}^{\frac{t^2}{2}}{\rm erf}\left(\frac{t}{\sqrt{2}}\right) \left(1-\epsb\,\frac{t^3}{3}\right)\right]\,, \label{eq:TPl2}\\
	Z_{-1}(s) &\sim	C^{(r)}_1 \, {\rm e}^{-\frac{s^2}{2}}\left(1+\epsb\,s-\epsb\,\frac{s^3}{3}\right) + C^{(r)}_2\left[\epsb\left(\frac{s^2}{3}-\frac{8}{3}\right)+\sqrt{\frac{\pi }{2}} {\rm e}^{-\frac{s^2}{2}} \text{erfi}\left(\frac{s}{\sqrt{2}}\right) \left(1+\epsb\,s-\epsb\,\frac{s^3}{3}\right)\right]\,.	\label{eq:TPr2}
\end{align}
We now match these expression that are valid around the turning points for $t,s\ll1$ to the WKB solutions valid far away from the turning points. To this end, we first express Eqs.~\eqref{eq:TPl2} and \eqref{eq:TPr2} in the limit $s,t\to\pm\infty$ using
\begin{align}
	\sqrt{\frac{\pi }{2}} t\,{\rm e}^{\frac{t^2}{2}}\text{erf}\left(\frac{t}{\sqrt{2}}\right)	&\sim	\pm\sqrt{\frac{\pi }{2}} e^{\frac{t^2}{2}} t-1+\frac{1}{t^2}-\frac{3}{t^4}\,,\qquad t\to\pm\infty\,,	\label{eq:matchTPl}\\
	\sqrt{\frac{\pi }{2}} e^{-\frac{s^2}{2}} \text{erfi}\left(\frac{s}{\sqrt{2}}\right)		&\sim	\frac{1}{s}+\frac{1}{s^3}\,,\qquad s\to\pm\infty\,.
\end{align}
This leads to
\begin{align}
	Z_{-1}(t) &\sim	\left(C^{(l)}_1\epsb\, {\rm e}^{-\frac1{6\epsm}}\pm C^{(l)}_2\sqrt{\frac{\pi }{2}}\right)t\,{\rm e}^{\frac{t^2}{2}}\left(1-\epsb\,\frac{t^3}{3}\right)+C^{(l)}_2\left[\frac{1}{t^2}-\frac{3}{t^4}+\epsb\,\frac{1}{t}-\epsb\,\frac{5}{t^3}\right]\,,\qquad t\to\pm\infty\,,\\
	Z_{-1}(s) &\sim	C^{(r)}_1 \, {\rm e}^{-\frac{s^2}{2}}\left(1+\epsb\,s-\epsb\,\frac{s^3}{3}\right) + C^{(r)}_2\left[\frac{1}{s^3}+\frac{1}{s}-2 \epsb\right]\,,\qquad s\to\pm\infty\,.
\end{align}
We now express the WKB solution \eqref{eq:WKB} in the matching variables $t$ and $s$ and expand to order $\epsb{}$:
\begin{align}
	Z^{\rm WKB}_{-1}(t)	&\sim	C_1\, \epsb {\rm e}^{-\frac{1}{6\epsm}} t\,{\rm e}^{\frac{t^2}{2}}\left(1-\epsb\,\frac{t^3}{3}\right) + \frac{C_2}{\epsb^2}\left[-\frac{1}{t^2}+\frac{3}{t^4}-\epsb\,\frac{1}{t}+\epsb\,\frac{5}{t^3}	\right]	\label{eq:matchWKBl}\,,\\
	Z^{\rm WKB}_{-1}(s)	&\sim	C_1\, {\rm e}^{-\frac{s^2}{2}}\left(1+\epsb\,s-\epsb\,\frac{s^3}{3}\right) + \frac{C_2}{\epsb}\left[ \frac{1}{s^3}+\frac{1}{s}-2 \epsb\right]\,.
\end{align}
It is now possible to determine the constants in front of the WKB solution in the different regions where the WKB approximation is valid. We call these constants $C^{(L)}_1$ and $C^{(L)}_2$ left of the turning point at $z=-1$, $C^{(M)}_1$ and $C^{(M)}_2$ between the two turning points and $C^{(R)}_1$ and $C^{(R)}_2$ right of the turning point at $z=0$. 

Since the solution is determined only up to a normalisation constant, we can choose $C^{(L)}_2=-1$ to obtain a positive power-law tail for large negative values of $z$. To ensure that the solution is normalisable, we must require $C^{(R)}_1=0$. Matching Eqs.~\eqref{eq:matchTPl} and \eqref{eq:matchWKBl} for $t<0$ then leads to $C^{(l)}_2=1/\epsm$ and $C^{(l)}_1=\sqrt{\frac{\pi}{2}}\epsb^{-3}{\rm e}^{1/(6\epsm)}$. Matching the same equation for $t>0$ we obtain the unknown coefficients $C^{(M)}_1$ and $C^{(M)}_2$, 
namely $C^{(M)}_1 = \sqrt{2\pi}\epsb^{-3}{\rm e}^{1/(6\epsm)}$ and $C^{(M)}_2=-1$. Matching through the turning point at $z=0$ is trivial because there the terms do not mix. We obtain $C^{(r)}_1=C^{(M)}_1=\sqrt{2\pi}\epsb^{-3}{\rm e}^{1/(6\epsm)}$ and $C^{(r)}_2=C^{(M)}_2=1$ and, furthermore $C^{(R)}_1=\sqrt{2\pi}\epsb^{-3}{\rm e}^{1/(6\epsm)}$ and $C^{(R)}_2=1$. Away from the turning points, the WKB approximation for $Z_{-1}$ now reads
\begin{widetext}
\begin{align}
\label{eq:wkb}
	Z^{\rm WKB}_{-1}(z) \sim \begin{cases}	-\frac{1}{z(z+1)^{2}}\Big[1+\epsm \big(\frac{1}{z^2 (z+1)^2}+\frac{4}{z (z+1)^2}\big)\Big]	&	z<-1	\\[0.1cm]
						\sqrt{2\pi}\epsb^{-3}{\rm e}^{1/(6\epsm)}(z+1)\exp[ -U(z)] - \frac{1}{z(z+1)^{2}}\Big[1+\epsm \big(\frac{1}{z^2 (z+1)^2}+\frac{4}{z (z+1)^2}\big)\Big]	&	-1<z<0	\\[0.1cm]
						\sqrt{2\pi}\epsb^{-3}{\rm e}^{1/(6\epsm)} (z+1)\exp[ -U(z)] + \frac{1}{z(z+1)^{2}}\Big[1+\epsm \big(\frac{1}{z^2 (z+1)^2}+\frac{4}{z (z+1)^2}\big)\Big] &z>0	\end{cases},
\end{align}
\end{widetext}
as $\eps\to0$.  Together with the approximations around the turning points, \eqref{eq:TPl2} and \eqref{eq:TPr2}, and the matching constants given above this 
constitutes a uniform approximation to $Z_{-1}(z)$ to order $\epsm$. We note that this solution is dominated by the exponentially term $\propto(z+1)\exp[-U(z)]$ in the regime $-1 \ll z \ll z^*$, where
\begin{align}
	z^* \sim \frac{1}{2}-\frac{4}{3} \epsm \log \left(\epsb^{3}\,\frac{8}{27} \sqrt{\frac{2}{\pi }} \right)
\end{align}
is the approximate zero of $Z_{-1}(z)$. In this region around $z=0$ we can simply neglect the power-law contribution. On the other hand, outside this regime 
the power-law tail dominates, while the exponential solution is either zero, or
it is strongly suppressed.

\section{Evaluation of the integral for $Z^{(1)}$}
We use the WKB approximation derived above to evaluate the integrals in Eq.~(\ref{eq:Z1}),
Eq.~(7) in the main text. We consider the positive tail of $Z^{(1)}(z)$ ($z\gg 1$). 
Substituting $t \to z - t$ and neglecting lower order terms in $z$ we obtain for $Z^{(1)}$:
\begin{align}
	Z^{(1)}(z)\!\sim\!\frac{z\!+\!1}{\epsm}\!\!
\int_{0}^{\infty}\!\!\!\!\!\!\ed t\,\frac{{\rm e}^{\frac1{\epsm}\left(-z^2 t-\frac{t^3}{3}+\frac{t^2}{2}\right)}}{(z-t+1)^2}\!\!\!\int_{-1}^{z-t}\!\!\!\!\ed t' t'(t'\!+\!1) Z_{-1}(t') \sim \frac{1}{\epsm z}\int_{0}^{\infty}\!\!\!\!\!\!\ed t\,{\rm e}^{\frac1{\epsm}\left(-z^2 t-\frac{t^3}{3}+\frac{t^2}{2}\right)}\!\!\!\int_{-1}^{z}\!\!\!\!\ed t' t'(t'\!+\!1) Z_{-1}(t')\,.
\end{align}
Note that we could neglect $t$ against $z$ because large $t$ are exponentially suppressed by the term $\propto{\rm e}^{-t^3/(3\epsm)}$.
For the inner integral substitute the turning point coordinate $s=t'/\epsb$ and expand to first order in \epsb{} to wit
\begin{align}
	\epsm\!\! \int_{-1/\epsb}^{z/\epsb}\!\!\!\!\ed s \,s(\epsb s\!+\!1) Z_{-1}(s)	&\sim	\sqrt{2\pi}\epsb^{-3}{\rm e}^{\frac1{6\epsm}}\int_{-1/\epsb}^{z^*\!\!/\epsb}\!\!\!\!\ed s\, {\rm e}^{-\frac{s^2}{2}}\left(1+\epsb\,s-\epsb\,\frac{s^3}{3}\right) - \int_{z^*}^{z}\!\!\!\!\ed t' \frac1{t'+1}\,,\\
			&\sim 2\pi {\rm e}^{\frac1{6\epsm}} -\log|z|\,, \qquad z\to\infty\,,
\end{align}
where the term arising from the boundary of the second term was dropped because it is negligible against the first constant. The remaining integral is obtained by performing integration by parts with respect to the $z$-dependent term. In this way
we find for the right tail:
\begin{align}
	Z^{(1)}(z) \sim \frac{2\pi {\rm e}^{\frac1{6\epsm}} -\log(z)}{\epsm z}\int_{0}^{\infty}\!\!\!\!\!\!\ed t\,{\rm e}^{\frac1{\epsm}\left(-z^2 t-\frac{t^3}{3}+\frac{t^2}{2}\right)}\sim 2\pi {\rm e}^{\frac1{6\epsm}} |z|^{-3} - |z|^{-3}\log |z|\,, \qquad z\to\infty\,.
\end{align}
A similar analysis for the left tail yields
\begin{equation}
Z^{(1)}(z) \sim  |z|^{-3}\log|z|\,, \qquad z\to-\infty\,,
\end{equation}
up to a negligible constant. The physical solution $Z_{\mu_{\rm c}}$ must have symmetric tails.
Noting that $Z_{-1}(z) \sim -z^{-3}$ for both tails, this condition gives
$ 1=          -1+ \delta\mu\,2\pi {\rm e}^{\frac1{6\epsm}}$. In other words:
\begin{align}
	\delta\mu &\sim {\pi}^{-1} {\rm e}^{-1/(6\epsm)}	\,.
\end{align}
This is Eq.~(8) in the main text.